\documentclass[12pt]{article}

\usepackage{url}  % Formatting web addresses
\usepackage[latin1]{inputenc}
\usepackage{amsmath}
\usepackage{amsfonts}

\newcommand{\Multi}{\text{Multi}}
\newcommand{\Dir}{\text{Dir}}
\newcommand{\Poi}{\text{Poisson}}
\renewcommand{\vec}[1]{\boldsymbol{#1}}

\newcommand{\expect}{\mathbb{E}}
\newcommand{\R}{\mathbb{R}}

\def\includegraphics{}

\begin{document}

\title{Simcluster: clustering enumeration gene expression data on the simplex space}

\author{Ricardo Z.N. Vêncio$^{1 \dag}$\footnote{to whom correspondence should be addressed: rvencio@gmail.com, \dag
contributed equally}, Leonardo Varuzza$^{2 \dag}$, \\Carlos A. de B. Pereira$^2$, Helena Brentani$^3$, Ilya
Shmulevich$^1$} \footnotetext[1]{Institute for Systems Biology, 1441 North 34th street, Seattle, WA 98103-8904, USA;
$^2$BIOINFO-USP - Núcleo de Pesquisas em Bioinformática, Universidade de São Paulo, São Paulo, Brazil; $^3$Hospital do
Câncer A. C. Camargo, São Paulo, Brazil}

\maketitle

\begin{abstract}

Transcript enumeration methods such as SAGE, MPSS, and sequencing-by-synthesis EST ``digital northern'', are important high-throughput techniques for
digital gene expression measurement. As other counting or voting processes, these measurements constitute compositional data exhibiting properties
particular to the simplex space where the summation of the components is constrained. These properties are not present on regular Euclidean spaces,
on which hybridization-based microarray data is often modeled. Therefore, pattern recognition methods commonly used for microarray data analysis may
be non-informative for the data generated by transcript enumeration techniques since they ignore certain fundamental properties of this space. Here
we present a software tool, Simcluster, designed to perform clustering analysis for data on the simplex space. We present Simcluster as a stand-alone
command-line C package and as a user-friendly on-line tool. Both versions are available at: http://xerad.systemsbiology.net/simcluster. Simcluster is
designed in accordance with a well-established mathematical framework for compositional data analysis, which provides principled procedures for
dealing with the simplex space, and is thus applicable in a number of contexts, including enumeration-based gene expression data.

\end{abstract}

%%%%%%%%%%%%%%%%
%% Background %%
%%
\section*{Background}

Technologies for high-throughput measurement of transcriptional gene expression are mainly divided into two categories: those based on hybridization,
such as all microarray-related technologies \cite{microarray,affymetrix} and those based on transcript enumeration, which include SAGE \cite{sage},
MPSS \cite{mpss}, and Digital Northern powered by traditional \cite{diginor} or, recently developed, EST sequencing-by-synthesis (SBS) technologies
\cite{ESTsolexa}.

Currently, transcript enumeration methods are relatively expensive and more time-consuming than methods based on hybridization. However, recent
improvements in sequencing technology, powered by the ``\$1000 genome'' effort \cite{1000genome}, promises to transform the transcript enumeration
approach into a fast and accessible alternative \cite{454,SBS,paper_helicos} paving the way for a systems-level absolute digital description of
individualized samples \cite{lhood}.

Methods for finding differentially expressed genes have been developed specifically in the context of enumeration-based techniques of different
sequencing scales such as EST \cite{audic}, SAGE \cite{rvencio_sagebetabin} and MPSS \cite{manosysbio}. However, in spite of their differences,
hybridization-based and enumeration-based data are typically analyzed using the same pattern recognition techniques, which are generally imported
from the microarray analysis field. 

In the case of clustering analysis of gene profiles, the simple appropriation of practices from the microarray analysis field has been shown to lead
to suboptimal performance \cite{poisson_sage}. Cai and co-workers \cite{poisson_sage} provided an elegant clustering computational solution to group
tag (rows in a usual expression matrix representation) profiles that takes into account the specificities of enumeration-based datasets.  However, to
the best of our knowledge, a solution for transcript enumeration libraries (columns in a usual expression matrix representation) is still needed. We
report on a novel computational solution, called Simcluster, to support clustering analysis of transcript enumeration libraries.

%%%%%%%%%%%%%%%%%%%%%%%%%%%%
%% Implementation %%
%%
\section*{Implementation}
  \subsection*{Theory}

Without loss of generality, we use the term ``tag'' to refer to the transcripts' representation, as usual in the SAGE field (this is equivalent to
the term ``signature'' in MPSS analysis or ``contigs'' in EST analysis). 

The theoretical model used here to describe the transcript enumeration process is the usual uniform sampling of interchangeable colored balls from an
infinite urn model. Given the total number $n$ of counted tags and the abundance vector $\vec \pi$ of all transcripts, this model leads to a
probabilistic description of the observed result: $\vec x|\vec \pi,n \sim \Multi(\vec \pi,n)$, i.e., the counts $\vec x$ follow a Multinomial
distribution \cite{rvencio_book}. It is also possible to model $\vec x$ as Poisson distributed \cite{poisson_model} since it is an approximation for
the Multinomial. Regardless of the specificities of the theoretical probabilistic model, it is well known that, as with other counting or voting
processes, the natural space for dealing with this kind of data is the simplex space.

The unitary simplex space, having $d$ dimensions, is defined as \cite{aitch_book, aitch_review}:

\begin{equation}
\label{eq:simplexdef}
S_{d-1} = \{ \vec \pi | \vec \pi \in \R_{+}^d, \vec \pi \vec 1' = 1 \}
\end{equation} 
where $\vec 1$ is a vector of ones. In the gene expression context, $d$ is the number of unique tags observed. 

An example of a simplex vector is $\vec p = \expect[\vec \pi|\vec x]$ and applying a standard Bayesian approach, one obtains from $\vec x|\vec
\pi,n$, using a Dirichlet prior density $\vec \pi \sim \Dir(\vec \alpha)$, the posterior density: $\vec \pi|\vec x \sim \Dir(\vec x + \vec \alpha)$.

It is known that clustering analysis is inherently dependent on the choice of a distance measure between the considered objects. This, in turn, is
connected to the structure of the underlying space. A metric $\Delta$, measuring the distance between two objects $a$ and $b$, must respect the
properties:

(i) $\Delta(a,b)=\Delta(b,a)$; 

(ii) $\Delta(a,b)=0\Leftrightarrow a=b$; 

(iii) $\Delta(a,c)\leq\Delta(a,b)+\Delta(b,c)$. 

One may also consider additional reasonable properties such as: 

(iv) scale invariance $\Delta(xa,yb)=\Delta(a,b),x,y \in \R_{+}$; and

(v) translational invariance $\Delta(a+t,b+t)=\Delta(a,b)$. 

These commonly required additional properties guarantee that distance measurements are not affected by the definition of arbitrary scale or
measurement units and that more importance is given to the actual difference between the objects being measured rather than commonalities.

Translations on the simplex space are defined by \cite{aitch_review}:

\begin{equation}
\label{eq:simplextranslation}
\vec p \oplus \vec t = \frac{(\vec p \cdot \vec t)}{(\vec p \cdot \vec t) \vec 1'}
\end{equation}
where $\cdot$ is the usual Hadamard product and the division is vector-evaluated.

Well known distances, such as Euclidean, Manhattan, and correlation-based distances, do not exhibit the properties (i)-(v) if the
measured objects belong to the simplex space, as is the case of transcript enumeration data. A possible metric that obeys (i)-(v) on the simplex 
space is the Aitchisonean distance \cite{aitch_review}:
\begin{equation}
\label{eq:aitch_dist}
\Delta(\vec p, \vec q) = \sqrt{ln\left( \frac{\vec p_{-d} / p_d}{ \vec q_{-d} / q_d} \right)
\left( \vec I + \vec 1' \times \vec 1 \right)^{-1}
ln\left( \frac{ \vec p_{-d}/p_d}{\vec q_{-d}/q_d} \right)'}
\end{equation} where $\vec I$ is the identity matrix, $\times$ is the Kronecker product, $-d$ subscript is a notation for ``excluding the $d^{th}$
element'', and elementary operations are vector-evaluated.

Clustering procedures coherent with this theoretical background are suitable for transcript enumeration data.

  \subsection*{Software design}

In short, Simcluster's method can be described as the use of a Bayesian inference step (currently with a uniform prior) to obtain the expected
abundance simplex vectors given the observed counts $\expect[\vec \pi|\vec x]$, and the use of the Aitchisonean distance in the following algorithms:
k-means, k-medoids and self-organizing maps (SOM) for partition clustering, PCA for inferring the number of variability sources present, and common
variants of agglomerative hierarchical clustering.

Currently, the Simcluster package is comprised of: Simtree, for hierarchical clustering; Simpart, for partition clustering; Simpca for Principal
Component Analysis (PCA); and several utilities such as TreeDraw, a program to draw hierarchical clustering dendrograms with user-defined colored
leaves. Simcluster's modularity allows relatively simple extension and addition of new modules or algorithms. Increasing the coverage of algorithms
and validity assessment methods \cite{bolshakova2005itm} are envisioned in future updates.

Simcluster can be used, modified and distributed under the terms of the GPL license \cite{GPLGNU}. The software was implemented in C for improved
performance and memory usage, assuring that even large datasets can be processed on a regular desktop PC.

To increase source code reuse, established libraries were used: Cluster 3 \cite{mhoon} for clustering, GNU Scientific library \cite{gsl} for PCA,
Cairo \cite{cairo} and a modification of TreeDraw X \cite{rpage} for colored dendrogram drawing.

The input data set can be a matrix of transcript counts or general simplex vectors. Some auxiliary shell and Perl scripts are available to:
automatically download data from the GEO database \cite{GEO}, convert GEO files to Simcluster input format, and filter out low-count tags.

The Linux-based installation and compilation is facilitated by a configuration script that detects all the prerequisites for Simcluster compilation.
Missing libraries are automatically downloaded from the Simcluster website and compiled by the Simcluster compilation process.

To broaden usability, a user-friendly web interface was developed and is made available at http://xerad.systemsbiology.net/simcluster$\_$web/. Figure 
\ref{fig:screenshot} shows a screenshot of an analysis session using Simcluster's 
web-based interface.

\begin{figure}[H]%
%\centerline{\includegraphics{fig01.pdf}}
\caption{Screenshot of an analysis session using Simcluster's web-based interface. Simcluster's on-line version was designed to be a user-friendly 
interface for the command-line version. The screenshot shown is an illustration of an interactive session usign the example data 
provided.}\label{fig:screenshot}
\end{figure} ***** FIGURE 01 HERE *****

%%%%%%%%%%%%%%%%%%%%%%%%%%%%
%% Results and Discussion %%
%%
\section*{Results and Discussion}

We agree with Dougherty and Brun \cite{probcluster, brun2007mbe} that ``validation'' of clustering results is a heuristic process, even though there
are some interesting efforts to objectively incorporate biological knowledge in this process using Gene Ontology, especially when one is clustering
gene expression profiles \cite{datta2006, loga06}. However, to illustrate the usefulness of our software, we collected several examples in which the
performance of Simcluster can be considered as qualitatively superior to some traditional approaches imported from the microarray analysis field.
These examples include EST, SAGE and MPSS datasets, and are available on the project's webpage \cite{simcluster}. Among these, we describe here a
simulated enumeration dataset built from real microarray data, for which we can define the ground truth and check results against it in a relatively
objective way. Of course, a comprehensive study with simulated data, consisting of comparisons of clustering algorithms, distance metrics, and
distributions generating the random point sets, would be necessary to properly evaluate any clustering algorithm. This should be the subject of
future work.

The objective of this example is to show that Simcluster is able to reconstruct the clustering result obtained for an Affymetrix microarray
dataset when the input is a simulated transcript enumeration dataset, built to mimic the real microarray biological data.

The data used to create the virtual transcript enumeration data was obtained from the Innate Immunity Systems Biology project \cite{innateimmunity}
and is provided as an Additional File. This data is a set of Affymetrix experiments of mouse macrophages stimulated by different Toll-like receptor
agonists (LPS, PIC, CPG, R848, PAM) during a time-course (0, 20, 40, 60, 80 and 120 minutes). A detailed description and biological significance of
this dataset is presented elsewhere \cite{sysbionature, innateimmunity}.

Using this data, a clustering analysis result is shown in Figure \ref{fig:affyarray}. This pattern is obtained using the most common type of
clustering analysis in the microarray field: Euclidean distance with average linkage agglomerative hierarchical clustering, implemented by R
\cite{Rwebsite} routines, available as Additional File. This clustering pattern will be considered to be the ``gold-standard'' for the purpose of
this simulation. 

\begin{figure}[H]%
%\centerline{\includegraphics{fig02.pdf}}
\caption{Clustering analysis of the Affymetrix dataset. Data produced by the Innate Immunity Systems Biology project \cite{sysbionature, 
innateimmunity}. This data is a set of Affymetrix experiments of mouse macrophages stimulated by different Toll-like receptor agonists (LPS, PIC, 
CPG, R848, PAM) during a time-course (0, 20, 40, 60, 80 and 120 minutes). Method: Euclidean distance with average linkage agglomerative hierarchical 
clustering.}\label{fig:affyarray}
\end{figure} ***** FIGURE 02 HERE *****

The virtual experiment consists of the creation of a transcriptome with the relative abundance between genes defined by the Affymetrix data; sampling
a random number of tags from it of different magnitudes; enumeration of sampled transcripts; and using some common clustering procedures along with
Simcluster.

It is easier to understand the concept of the virtual transcriptome by following a particular case. For the sample labeled LPS-120 measured 120
minutes after the LPS stimulus, the Affymetrix expression levels are:

\begin{tabular}{cccc}
\\
\end{tabular}

\begin{tabular}{llll}
Probesets & Representative ID & Gene Symbol & Intensity (sorted) \\
\hline
1457375\_at  & BG094499 & Transcribed locus & 1.94760 \\
1452109\_at  & BG973910 & interleukin 17 receptor E & 2.14522 \\
$\cdots$    & $\cdots$  & $\cdots$ & $\cdots$  \\
M12481\_3\_at & AFFX-b-ActinMur & actin beta cytoplasmic & 36191.41765 \\
1436996\_x\_at  & AV066625 & P lysozyme structural & 43458.17590 \\
\end{tabular}

\begin{tabular}{cccc}
\\
\end{tabular}

The virtual total number of available tags is defined as proportional to the measured intensity using 10,000 as a scaling constant, an arbitrary
number large enough to assure that finite population issues are negligible. Actual examples are: 19,476 for BG094499; 21,452 for BG973910; and so on
until 361,914,176 for actin; and 434,581,759 for AV066625. The total amount of available tags is $T$ = 126,971,909,452, which is a number much
greater than the typical number of sequenced tags and is in accordance with the ``infinite urn'' model.

The total of virtually sequenced tags $N$ for each sample is simulated from a Poisson distribution, $N \sim \Poi(n)$, to create a realistic virtual
sequencing library. All generated data and results are available as Additional Files. For example, the actual simulation for $n$ = 1,000,000
virtually sequenced tags assigned $N$ = 1,001,794 for the LPS-120 library; $N$ = 998,382 for the CPG-40 library; and so on. The same process is
repeated for increasing $n$ from 100,000 to 100,000,000. Since $n \ll T$ for all $n$ considered, the multinomial sampling is used and its mean is
taken for each library, according to the assumed ``infinite urn'' model. The results for the largest simulation are shown in Figures
\ref{fig:lastsimulaffy1}-\ref{fig:lastsimulaffy4} and individual results for all separate increasing $n$ sizes are available as Additional Files.

\begin{figure}[H]%
%\centerline{\includegraphics{fig03.pdf}}
\caption{Simcluster's clustering of simulated data based on Affymetrix expression levels. Transcript enumeration data produced by the simulation of a 
virtual transcriptome according to the Affymetrix expression levels. Sample size $n$ = 100,000,000. Method: Simcluster's average linkage 
agglomerative hierarchical clustering.}\label{fig:lastsimulaffy1}
\end{figure} ***** FIGURE 03 HERE *****

\begin{figure}[H]%
%\centerline{\includegraphics{fig04.pdf}}
\caption{Clustering of simulated data using Euclidean distance. Transcript enumeration data produced by the simulation of a virtual transcriptome 
according to the Affymetrix expression levels. Sample size $n$ = 100,000,000. Method: Euclidean distance with average linkage agglomerative 
hierarchical clustering.}\label{fig:lastsimulaffy2}
\end{figure} ***** FIGURE 04 HERE *****

\begin{figure}[H]%
%\centerline{\includegraphics{fig05.pdf}}
\caption{Clustering of simulated data using correlation distance. Transcript enumeration data produced by the simulation of a virtual transcriptome 
according to the Affymetrix expression levels. Sample size $n$ = 100,000,000. Method: correlation-based distance with average linkage agglomerative 
hierarchical clustering.}\label{fig:lastsimulaffy3}
\end{figure} ***** FIGURE 05 HERE *****

\begin{figure}[H]%
%\centerline{\includegraphics{fig06.pdf}}
\caption{Clustering of simulated data using cosine distance. Transcript enumeration data produced by the simulation of a virtual transcriptome 
according to the Affymetrix expression levels. Sample size $n$ = 100,000,000. Method: cosine distance with average linkage agglomerative hierarchical 
clustering.}\label{fig:lastsimulaffy4}
\end{figure} ***** FIGURE 06 HERE *****

It is clear that cluster results obtained by Simcluster converge to the same structure obtained by analyzing the Affimetrix data, as the number of
virtually sequenced tags increases. Moreover, Simcluster's results are not only compatible with the usual microarray analysis for Affymetrix data,
but also are more biologically meaningful than the results obtained by the usual microarray analysis techniques applied to the virtual sequencing
data.  As in the original microarray analysis, the Simcluster result is able to cluster together the different stimuli, placing consecutive
time-points close to each other. 

Although this kind of analysis certainly does not provide a proof, the above result indicate that the theoretical framework is adequate for
enumeration-based data, as expected. Additional examples and discussions can be found on the project's website \cite{simcluster}.

%%%%%%%%%%%%%%%%%%%%%%
\section*{Conclusions}

We developed a software tool, called Simcluster, for clustering libraries of enumeration-based data. It is important to note that Simcluster is built
in accordance with a well-established mathematical framework for compositional data analysis, which provides principled procedures for dealing with
the simplex space, and is thus applicable in contexts other than transcript enumeration.

\section*{Authors contributions}

RZNV proposed and conducted the study. LV wrote the software and helped to interpret the results. CABP indicated the compositional analysis
literature and is LV's PhD thesis advisor. HB provided biological insight for result interpretation. IS supervised the study. RZNV and IS wrote the
manuscript. All authors read and approved the final manuscript.

\section*{Acknowledgements}

We thank Dr. Jared Roach (ISB) and Dr João C. Barata (USP) for constructive discussions and Dr. Alistair Rust (ISB) for help with the web server. LV
is supported by a CAPES. CABP is partially supported by CNPq. This work is partially supported by NIH/NIAID grants U19-AI057266 and U54-AI54253 and
NIH/NIGMS P50-GMO-76547.

%%%%%%%%%%%

\bibliography{bmc_article}

%%%%%%%%%%%

\end{document}